# Theoretical and experimental investigation on tandem modulator configurations for Frequency Coded Quantum Key Distribution systems

J. Mora, A. Ruíz-Alba, W. Amaya, V. García-Muñoz, J. Capmany

*Optical and Quantum Communications Group, Institute of Telecommunications and Multimedia (ITEAM), Universidad Politécnica de Valencia, Edificio 8G, Camino de Vera s/n, 46021 Valencia, SPAIN*

[*]*Corresponding author: jcapmany@iteam.upv.es*

We have theoretically and experimentally address the possible tandem electro-optical modulator configurations that can be employed to implement Frequency Coded Quantum Key Distribution systems (FC-QKD). A closed and general formulation of the end to end field propagation in a dispersion compensated FC-QKD optical fiber system under the low modulation index regime is presented which accounts for all the possible tandem combinations. The properties and the parameter selection of the modulators to achieve each one are summarized. We also address which protocol (B92, BB84 or either) is feasible to be implemented with each configuration. The results confirm those reported for the configurations previously reported in the literature and, at the same time, show the existence of four novel tandem modulator configurations that can also be employed. We have also provided experimental evidence of the successful operation of the novel configurations that confirm the behavior predicted by the theoretical results.

*OCIS codes: 270.5565 Quantum Communications, 130.411 Modulators, 060.5060 Phase modulation, 230.2090 Electro-optical devices.*

# 1. Introduction

Quantum cryptography features an unique way of sharing a random sequence of bits between users with a certifiable security not attainable with either public or secret-key classical cryptographic systems [1],[2]. This is achieved by means of quantum key distribution (QKD) techniques.

Photonics has proved to be one of the principal enabling technologies for long-distance QKD using optical fiber links. One approach of particular interest is the so called frequency coding FC-QKD, which relies on encoding the information bits on the sidebands of either phase [3] or amplitude [4] radiofrequency (RF) modulated light. In FC-QKD systems, as shown in figure 1, Alice (the emitter) randomly changes the phase of the electrical signal used to drive a light modulator among four phase values $0, \pi$ and $\pi/2, 3\pi/2$, which form a pair of conjugated bases. When it arrives at Bob (the receiver), he modulates the signal again using a second external modulator with the same microwave signal frequency and thus his new sidebands will interfere with those created by Alice. A subsequent filtering stage helps to detect the presence of photons in the upper or lower modulation sidebands upon which a decision is made on whether a "0" or a "1" has been received. The capacity of FC-QKD systems can be increased, by using several RF subcarriers both at Alice and Bob's modulators leading to subcarrier multiplexed SCM-QKD systems [5].

Both FC and SCM QKD systems require the use of a pair or tandem of modulators. In principle, there are three types of available devices in the market, which are the phase modulator (PM), the amplitude modulator (AM) and the unbalanced modulator (UM). The phase modulator only varies the phase of the propagating field, the AM modulator only changes the signal amplitude and is chirp-free and the unbalanced modulator (UM) changes both the phase and the amplitude.

Several combinations have been reported in the literature for the implementation of frequency coded QKD. For instance, PM-PM [6] and AM-AM [7] configurations have been proposed for QKD systems implementing the B92 [2] protocol, while the UM-UM [4] and AM-PM [7] configurations have been reported for QKD systems implementing the BB84 [2] protocol. Yet however, there are other possible configurations that have not been yet considered and could lead to successful operation either for B92 or BB84 protocols. The purpose of this paper is precisely to explore these novel configurations and also to provide a theoretical framework that unifies the results for all the possible configurations.

The paper is structured as follows. We first of all present a closed and general formulation of the end to end field propagation in a dispersion compensated FC-QKD optical fiber system under the low modulation index regime. This formulation is general and accounts for all the possible tandem combinations. The properties and the parameter selection of the modulators to achieve each one are summarized in a reference table. We also illustrate which protocol (B92, BB84 or either) is feasible to be implemented with each configuration. The results confirm those reported for the configurations previously reported in the literature and, at the same time, show the existence of four novel tandem modulator configurations which can also be employed. In section 3 we provide the experimental evidence of the successful operation of the novel configurations that confirm the behavior predicted by the theoretical results.

The theoretical and experimental models only consider the case of frequency coded or subcarrier multiplexed QKD systems based on or faint pulse sources, where the output signal from Alice can be represented by a coherent state and thus the system performance description and evaluation can be carried under classical regime in order to check the principle of operation.

## 2. General Formalism for Tandem Modulator FC-QKD Systems

Frequency coded quantum key distribution systems consist of a monochromatic optical source that is externally modulated by a phase modulated RF signal of frequency Ω at Alice's premises [2],[4], an optical fiber link and finally at the receiver or Bob's side the light is again externally modulated by another phase modulated RF signal of the same frequency Ω as that employed at Alice's transmitter and detected after an optical filtering process. Figure 1 illustrates the scheme which we emphasize, is based on the cascade of two modulators.

### *2.1 End to End Optical field propagation*

We now present a theoretical that accounts for the aforementioned system regardless of which type modulator (PM, AM and UM) is employed at Alice or Bob's location. The final expression is thus more general and complete that any other previously reported in the literature and will allow us to explore the viability of previously unreported cascade configurations.

First of all, we consider that the laser source emits a monochromatic wave with frequency $\omega_0$ given by:

$$E_0(t) = E_0 e^{-j\omega_o t} \qquad (1)$$

where $E_0$ is the complex amplitude, the field at the output of Alice location can be expressed as

$$E_{ALICE}(t) = E_0 e^{-j\omega_o t} \left\{ \varepsilon_{1A} e^{j\psi_A} \cdot e^{jm_{1A} \cdot cos(\Omega t + \Phi_A)} + \varepsilon_{2A} e^{-j\psi_A} \cdot e^{-jm_{2A} \cdot cos(\Omega t + \Phi_A)} \right\} \qquad (2)$$

Here $\psi_A$ is related to the bias voltage applied to Alice's modulator and $\varepsilon_{iA}$ is the coupling factor with $i=1$ for the upper arm and $i=2$ for the lower arm of a Mach Zehnder interferometer (MZI). The term $\Phi_A$ represents the electrical phase that Alice can choose ($\Phi_A = 0, \pi/2, \pi, 3\pi/2$) and $m_{iA}$ is the modulation index that initially we take different for the upper arm (i=1) and for the lower arm (i=2) of a Mach Zehnder interferometer (MZI).

The reader should note that with equation (2) we can describe the behavior of any of the three possible modulators just by properly selecting the value of the coefficients $m_{iA}$ and $\varepsilon_{iA}$. This fact allows us to have a closed and compact expression for all the possible configurations. For example, the AM modulator is characterized by the fact that the two coupling factors and the two modulation indexes are equal for both arms. For the case of the unbalanced modulator, the coupling factors are equal, while one of the modulation indexes has to be null, for example, for the lower arm $m_{2A}=0$, and the phase difference between the two arms is $2\psi_A$. Finally, for the phase modulator we have only one arm which can be represented as a MZI with zero coupling in the second arm ($\varepsilon_{2A}=0$). In table I we summarize the values corresponding to each configuration. In practice, all frequency coded and subcarrier multiplexed quantum key distribution systems are operated under the so called low modulation regime ($m_{iA} \ll 1$) in order to avoid intermodulation effects [8]. In this case the field at the output of Alice's modulator can be expressed as

$$E_{ALICE}(t) = E_0 e^{-j\omega_o t} \left\{ E_A(\omega_o) + E_A(\omega_o - \Omega)e^{j\Omega t} + E_A(\omega_o + \Omega)e^{-j\Omega t} \right\} \quad (3)$$

where $E_A(\omega_o)$ and $E_A(\omega_o \pm \Omega)$ represent the amplitude of the optical carrier $\omega_o$ and both sidebands respectively.

$$E_A(\omega_o) = \varepsilon_{1A} e^{j\psi_A} + \varepsilon_{2A} e^{-j\psi_A}$$
$$E_A(\omega_o \pm \Omega) = j\frac{1}{2}\left(\varepsilon_{1A} m_{1A} \cdot e^{j\psi_A} - \varepsilon_{2A} m_{2A} \cdot e^{-j\psi_A}\right) \cdot e^{\mp j\Phi_A} \quad (4)$$

The field propagates through the fiber link with length *L* and arrives to Bob's modulator. If we consider that the dispersion effects are compensated, the field at Bob's side, prior to the external modulation, is

$$E_S(t) = E_0 e^{-j\omega_o(t-\beta_1 L)} \cdot \left\{ E_A(\omega_o) + E_A(\omega_o - \Omega)e^{j\Omega(t-\beta_1 L)} + E_A(\omega_o + \Omega)e^{-j\Omega(t-\beta_1 L)} \right\} \quad (5)$$

where $\beta_1$ is the propagation constant at the optical frequency $\omega_o$ which is related to the optical delay of all bands after propagation.

At the output of Bob's modulator, the field is

$$E_{BOB}(t) = E_S(t) \cdot \left\{ E_B(\omega_o) + E_B(\omega_o - \Omega)e^{j\Omega t} + E_B(\omega_o + \Omega)e^{-j\Omega t} \right\} \quad (6)$$

The terms $E_B(\omega_o)$ and $E_B(\omega_o \pm \Omega)$ inside the bracket correspond to the amplitude of the optical carrier and subcarrier sidebands generated by Bob's modulator if Alice's modulator is not present. Therefore, they are given by similar expressions to those corresponding to Alice's modulator (4). Taking this fact into account the field $E_{BOB}(t)$ can be finally expressed as

$$E_{BOB}(t) = E_{BOB}(\omega_o) + E_{BOB}(\omega_o - \Omega)e^{j\Omega t} + E_{BOB}(\omega_o + \Omega)e^{-j\Omega t} \quad (7)$$

where we have neglected the harmonic distortion terms since the modulation index is low, and:

$$\begin{aligned} E_{BOB}(\omega_o) &= E_A(\omega_o) E_B(\omega_o) \\ E_{BOB}(\omega_o - \Omega) &= E_B(\omega_o) E_A(\omega_o - \Omega) \cdot e^{-j\Omega \beta_1 L} + E_A(\omega_o) E_B(\omega_o - \Omega) \\ E_{BOB}(\omega_o + \Omega) &= E_B(\omega_o) E_A(\omega_o + \Omega) \cdot e^{j\Omega \beta_1 L} + E_A(\omega_o) E_B(\omega_o + \Omega) \end{aligned} \quad (8)$$

This signal is the subject to the effect of two filters. One selects the content of the sideband centered at $\omega_o + \Omega$ (upper sideband) and the other selects the content of the sideband centered at $\omega_o - \Omega$ (lower sideband). The output of each filter is then sent to a different photon counter. The received power in the upper and lower sideband photon counters can be expressed in a closed and compact form that embraces any possible combination of Alice-Bob modulator cascade.

$$P(\omega_o \pm \Omega) = \frac{1}{2}\left[1 + V \cdot \cos\left(\Phi_B - \Phi_A + \Omega\beta_1 L \pm \Theta\right)\right] \tag{9}$$

The visibility of each band is:

$$V = \frac{2|\kappa_o||\kappa_1|}{|\kappa_o|^2 + |\kappa_1|^2} \tag{10}$$

And the parameters $\kappa_o$ and $\kappa_1$ are defined as:

$$\begin{aligned}\kappa_o &= \left(\varepsilon_{1B}e^{j\psi_B} + \varepsilon_{2B}e^{-j\psi_B}\right)\cdot\left(\varepsilon_{1A}m_{1A}\cdot e^{j\psi_A} - \varepsilon_{2A}m_{2A}\cdot e^{-j\psi_A}\right)\\ \kappa_1 &= \left(\varepsilon_{1A}e^{j\psi_A} + \varepsilon_{2A}e^{-j\psi_A}\right)\cdot\left(\varepsilon_{1B}m_{1B}\cdot e^{j\psi_B} - \varepsilon_{2B}m_{2B}\cdot e^{-j\psi_B}\right)\end{aligned} \tag{11}$$

With $\Theta$ being the phase difference between $\kappa_o$ and $\kappa_1$.

## 2.2 Possible protocol implementations using tandem modulator configurations

The formalism developed in section 2.1 is completely general and has the added value of allowing us to check which QKD protocol, either BB84 or B92 can be supported by any given modulator cascade configuration. We first start by developing the general conditions that must be fulfilled by a frequency coded or subcarrier multiplexed QKD system in order to implement the B92 or BB84 protocols.

The BB84 protocol is implemented in a frequency coded QKD system when the normalized powers of the side bands verify:

$$P(\omega_o + \Omega) = \cos^2(\Delta\Phi/2)$$
$$P(\omega_o - \Omega) = \sin^2(\Delta\Phi/2) \quad (12)$$

The above relationships, when translated to the general expressions of eq. (9) require two conditions for the implementation of this protocol. The first one is affects the visibility V of the system while the second affects to the phase difference $\Theta$:

$$V = 1 \Leftrightarrow |\kappa_o| = |\kappa_1|$$
$$\Theta = (2m+1)\frac{\pi}{2} \quad m = 0, \pm 1, \pm 2... \quad (13)$$

As far as the B92 protocol is concerned, its implementation requires the following condition on the normalized powers of the sidebands:

$$P(\omega_o \pm \Omega) = \cos^2(\Delta\Phi/2) \quad (14)$$

In this case, two conditions must be fulfilled:

$$V = 1 \Leftrightarrow |\kappa_o| = |\kappa_1|$$
$$\Theta = n\pi \quad n = 0, \pm 1, \pm 2... \quad (15)$$

For a given modulator cascade configuration, when the two conditions are fulfilled simultaneously either for the BB84 protocol (eq. 13) or for the B92 protocol (eq. 15) their implementation is carried by adjusting the phases $\Phi_A$ and $\Phi_B$ to provide the necessary required values for $\Delta\Phi = 0$, $\pi/2$, $\pi$ and $3\pi/2$. This is done, by taking into account that the term $\Delta\Phi$ follows the expression:

$$\Delta\Phi = \Phi_B - \Phi_A + \Omega\beta_1 L + \Theta \quad (16)$$

At this point, and once we have established the general conditions that any tandem modulator combination must fulfill to implement one of the mentioned protocols, we will describe the possible setups for each of them.

In table 2 we summarize, for each possible tandem modulator combination, the feasibility of implementing either the B92 or the BB84 protocol and the required values of their relevant parameters as referred to equation (9). The reader can identify in the table several configurations that have been previously reported in the literature (the proper reference appears in the last column when applicable). In particular, the configuration UM-UM has been the most commonly employed in the past to implement the BB84 protocol [4] although it can be used as well for implementing the B92 protocol by applying the same bias voltage to Alice and Bob's modulators. Indeed, this is the only configuration that allows the implementation of both protocols as shown in table 2. The rest of combinations only allow either the implementation of the B92 or BB84 protocol. For instance, the AM-AM and PM-PM only allow the implementation of the B92 protocol.

In table 2 we can also observe four configurations that, up to our knowledge, have not been previously reported in the literature while being feasible for implementing either the BB84 or the B92 protocol. In the first pair of configurations (UM-PM and PM-UM), only the B92 can be implemented while the last configurations that combines UM and AM only the BB84 protocol is possible, as we will see.

For the UM-PM set up, the $\kappa$ factors are:

$$\kappa_o = \frac{1}{2} m_A e^{j(\psi_A + \psi_B)}$$
$$\kappa_1 = m_B \cos(\psi_A) e^{j\psi_B}$$
(17)

In this case, if the protocol imposes that the UM modulator bias $\psi_A$ is an even multiple of $\pi/2$, the unitary visibility condition is not reached since the factor $\kappa_1$ vanishes. Therefore, the BB84 can not be implemented since it requires that $\psi_A=(2n+1)\pi/2$ while the B92 is possible since the condition $\psi_A=n\pi$ is compatible with the condition $V=1$. For the PM-UM configuration, the same conditions are found for the UM modulator.

For the UM-AM setup, the κ factors are:

$$\kappa_o = \frac{1}{2} m_A \cos(\psi_B) e^{j\psi_A}$$
$$\kappa_1 = m_B \cos(\psi_A) \sin(\psi_B) e^{j\pi/2}$$
(18)

The B92 condition shown in eq. (19) imposes that the phase of the UM modulator $\psi_A$ is $\pi/2$ but, in this case, the factor $\kappa_o$ is null and no photon is received at the counters (the visibility is null). Thus, the B92 protocol can not be implemented using this configuration. As far as the implementation of the BB84 protocol is concerned, $\psi_A$ must be a multiple of $\pi$, which corresponds to a maximum or minimum of the modulator optical power transfer function. This condition is perfectly compatible with the requisite for unitary visibility and consequently, the BB84 protocol can be implemented with this configuration. For the AM-UM configuration, the same conditions are found for the UM modulator.

## 3. Experimental results

In this section, we provide the experimental results confirming the performance of the novel structures as predicted by the general theoretical analysis presented in section 2.

Figure 2 illustrates the experimental setup that is used to validate the theoretical results for the novel reported schemes UM-PM and UM-AM. The optical source is a CW laser operating at 1550 nm featuring an optical power of 5 dBm and a linewidth of 200 MHz. We have employed three different modulators with an electrical bandwidth around 12.5, 20 and 30 GHz for the AM, PM and UM, respectively. Regarding to the half-wave voltage, each modulator has a value around 4.7, 7.4 and 5.5 V, respectively. In the AM and UM modulators the bias voltage is supplied by a DC voltage source which determines the operation point of each modulator according to the DC phases $\psi_A$ and $\psi_B$. Two local oscillators (OL) were used to drive the modulators of Alice and Bob with a frequency of modulation of 15 GHz. Each one of the modulators A and B incorporates an electrical attenuator and a phase shifter in order to control independently the modulation index ($m_A$ and $m_B$) and the variable phase difference through the electrical phase of each modulator, i.e., $\Phi_A$ and $\Phi_B$.

As mentioned earlier in the paper we only consider faint pulse frequency coded or subcarrier multiplexed QKD systems where the output signal from Alice can be represented by a coherent state. Thus the test of the system performance can be carried under classical regime in order to check the principle of operation. The process of detection is realized by means an optical spectrum analyzer (OSA) with a spectral resolution of 10 pm.

The first checked setup is the UM-PM configuration where Alice's modulator is UM and Bob's modulator is PM. Taking into account the considerations of previous section, we confirm that only the protocol B92 is possible when $\psi_A=0$ (i.e. the UM bias voltage is null). Figure 3 (a) shows the measured evolution of the optical power of each side band ($\omega_o - \Omega$ and $\omega_o + \Omega$) with the phase difference between Alice and Bob ($\Delta\Phi$). We observe a good agreement with the experimental points and the theoretical prediction (solid line). In Figure 3 (b) we can see the power evolution measurements for the complementary combination (PM-UM). We observe that the behavior is identical in both structures UM-PM and PM-UM as expected. In order to demonstrate the B92 protocol implementation, in Figure 4 we show the optical spectra measured at the output of the system for three relative phase differences $\Delta\Phi$: 0, $\pi/2$ and $\pi$. In all the plots, we observe a central peak that corresponds to the optical carrier. In Figure 4 (a) and (b) we observe two side bands separated 15 GHz from the carrier. The sidebands are present when $\Delta\Phi$ is 0 or $\pi/2$ and disappear when $\Delta\Phi=\pi$, demonstrating the implementation of the B92 frequency coded protocol.

In second place, we analyze the schemes working with the two possible combinations including an AM and an UM. In both cases, the UM modulator DC phase ($\psi_A$ or $\psi_B$) must be an even multiple of $\pi$. Therefore, we do not apply any bias voltage to the UM. Figure 5 plots the experimental evolution of the sidebands optical power with the phase difference $\Delta\Phi$ for both combinations (AM-UM and UM-AM). We can observe that the experimental dependence of both sidebands with $\Delta\Phi$ corresponds with the protocol BB84 as the equation (12) shows. Additionally, we check again the symmetry with the AM-UM and UM-AM schemes when the initial phase is adjusted to satisfy the expression (16). Finally, Figure 6 illustrates the optical spectra measured at the output of the system when the phase difference $\Delta\Phi$ is 0, $\pi/2$ and $\pi$, demonstrating the implementation of the BB84 protocol.

## 4. Summary and Conclusions

In this paper we have theoretically and experimentally addressed the possible tandem electro-optical modulator configurations that can be employed to implement Frequency Coded Quantum Key Distribution systems (FC-QKD). We have presented a closed and general formulation of the end to end field propagation in a dispersion compensated FC-QKD optical fiber system under the low modulation index regime. This formulation is general and accounts for all the possible tandem combinations. The properties and the parameter selection of the modulators to achieve each one have been summarized. We also addressed which protocol (B92, BB84 or either) is feasible to be implemented with each configuration. The results confirm those reported for the configurations previously reported in the literature and, at the same time, show the existence of four novel tandem modulator configurations that can also be employed. We have also provided experimental evidence of the successful operation of the novel configurations that confirm the behavior predicted by the theoretical results.

## Acknowledgments

The authors wish to acknowledge the financial support of the Spanish Government through Quantum Optical Information Technology (QOIT), a CONSOLIDER-INGENIO 2010 Project and the Generalitat Valenciana through the PROMETEO research excellency award programme GVA PROMETEO 2008/092.

# TABLE CAPTIONS

TABLE 1: Coefficient values for each possible modulator implementation.

TABLE 2: Parameter requirements for the different tandem modulator configurations to implement the BB84 and the B92 protocols in a frequency coded QKD system using a dispersion compensated link.

# FIGURE CAPTIONS

Figure 1: Diagram of a frequency coded QKD system employing a tandem modulator configuration.

Figure 2. Experimental setup operating in the classical regime.

Figure 3. Lower sideband (●), upper sideband (■) and theoretical power (solid line) versus $\Delta\phi$ when: (a) Alice operates with a PM and Bob has an UM and (b) vice-versa.

Figure 4. Spectra for the PM-UM configuration for different phase differences $\Delta\Phi$: a) 0, b) $\pi/2$ and c) $\pi$. The frequency is gauged to the carrier frequency.

Figure 5. Lower sideband (●), upper sideband (■) and theoretical power (solid line) versus $\Delta\phi$ when: (a) Alice operates with a AM and Bob has an UM and (b) vice-versa.

Figure 6. Spectra for the AM-UM configuration for different phase differences $\Delta\Phi$: a) 0, b) $\pi/2$ and c) $\pi$. The frequency is gauged to the carrier frequency.

**Tables**

|     | m | ε |
|-----|---|---|
| PM  | $m_{1A} = m_A$ ; $m_{2A} = 0$ | $\varepsilon_{1A} = \varepsilon_A$ ; $\varepsilon_{2A} = 0$ |
| AM  | $m_{1A} = m_{2A} = m_A$ | $\varepsilon_{1A} = \varepsilon_{2A} = \varepsilon_A$ |
| UM  | $m_{1A} = m_A$ ; $m_{2A} = 0$ | $\varepsilon_{1A} = \varepsilon_{2A} = \varepsilon_A$ |

**TABLE 1**

| A | B | Θ | V=1 | B92 | BB84 | REF |
|---|---|---|---|---|---|---|
| UM | UM | $\psi_B - \psi_A$ | $\dfrac{m_A}{m_B} = \dfrac{|\cos(\psi_A)|}{|\cos(\psi_B)|}$ | OK<br>$\psi_B = \psi_A + n\pi$ | OK<br>$\psi_B = \psi_A + (2n+1)\pi/2$ | [4] |
| AM | AM | 0 | $\dfrac{m_A}{m_B} = \dfrac{|\tan(\psi_B)|}{|\tan(\psi_A)|}$ | OK | NO | [7] |
| PM | PM | 0 | $\dfrac{m_A}{m_B} = 1$ | OK | NO | [6] |
| PM | AM | $\dfrac{\pi}{2}$ | $\dfrac{m_A}{m_B} = |\tan(\psi_B)|$ | NO | OK | [7] |
| AM | PM | $-\dfrac{\pi}{2}$ | $\dfrac{m_A}{m_B} = \dfrac{1}{|\tan(\psi_A)|}$ | NO | OK | [7] |
| UM | PM | $-\psi_A$ | $\dfrac{m_A}{m_B} = 2|\cos(\psi_A)|$ | OK<br>$\psi_A = n\pi$ | NO<br>V = 0 if<br>$\psi_A = (2n+1)\pi/2$ | NEW |
| PM | UM | $\psi_B$ | $\dfrac{m_A}{m_B} = \dfrac{1}{2|\cos(\psi_B)|}$ | OK<br>$\psi_B = n\pi$ | NO<br>V = 0 if<br>$\psi_B = (2n+1)\pi/2$ | NEW |
| UM | AM | $\dfrac{\pi}{2} - \psi_A$ | $\dfrac{m_A}{m_B} = 2|\cos(\psi_A)\tan(\psi_B)|$ | NO<br>V = 0 if<br>$\psi_A = (2n+1)\pi/2$ | OK<br>$\psi_A = n\pi$ | NEW |
| AM | UM | $-\dfrac{\pi}{2} + \psi_B$ | $\dfrac{m_A}{m_B} = 2\dfrac{|\cos(\psi_B)|}{|\tan(\psi_A)|}$ | NO<br>V = 0 if<br>$\psi_B = (2n+1)\pi/2$ | OK<br>$\psi_B = n\pi$ | NEW |

**TABLE 2**

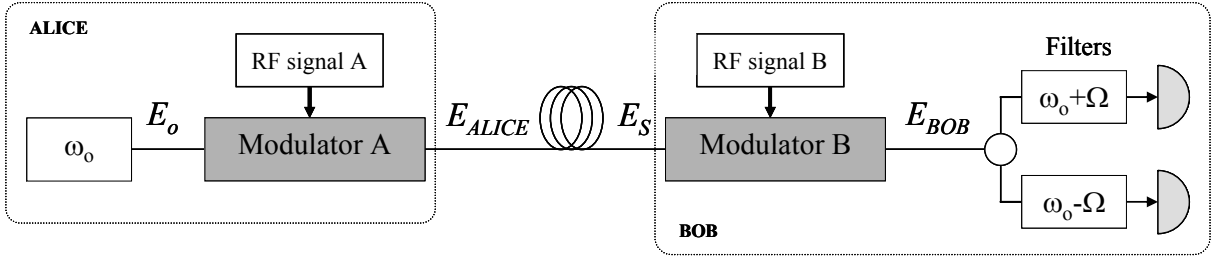

**FIGURE 1**

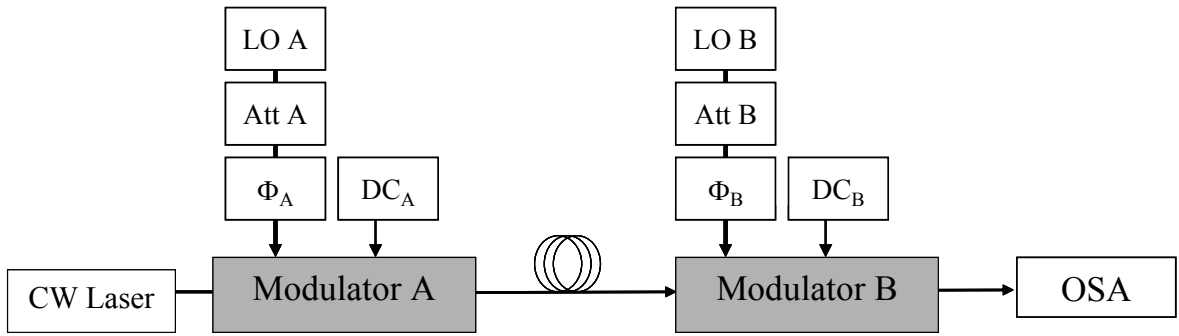

**FIGURE 2**

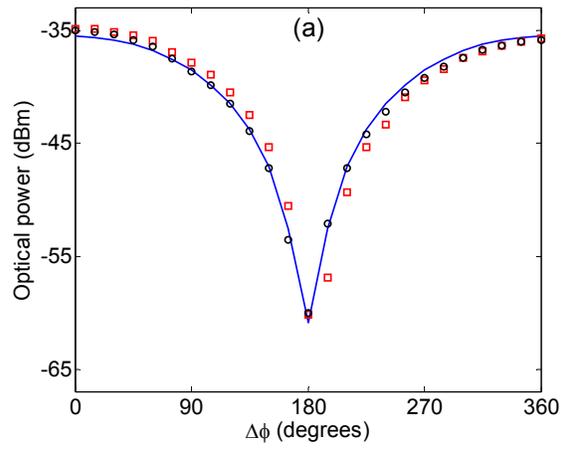

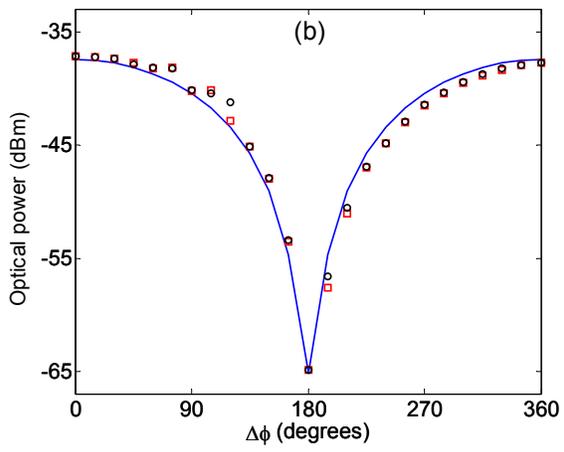

**FIGURE 3**

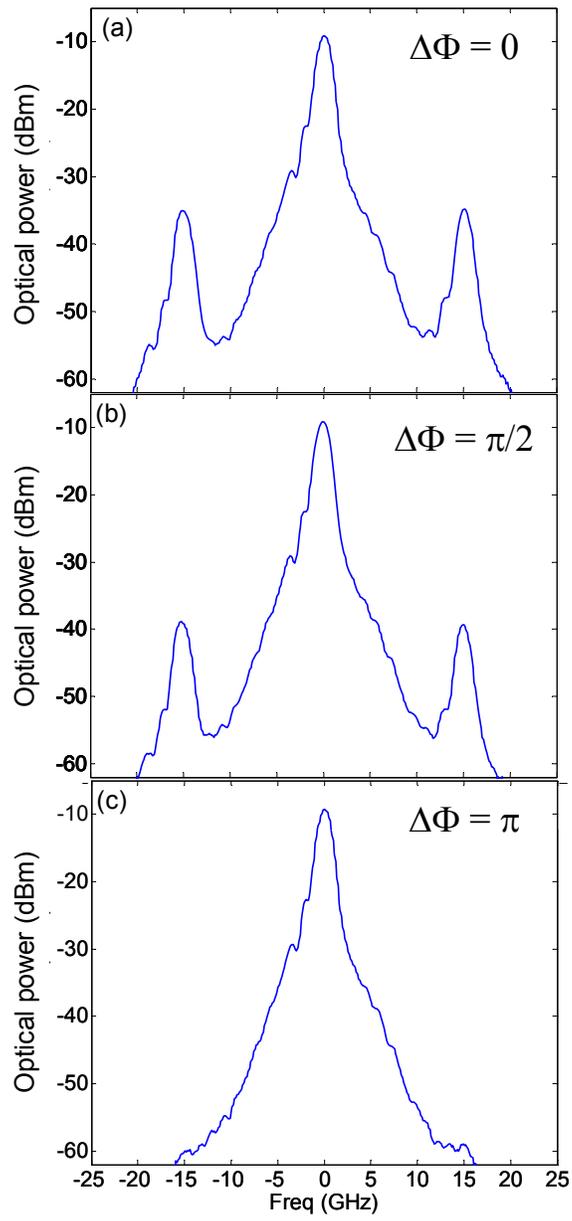

**FIGURE 4**

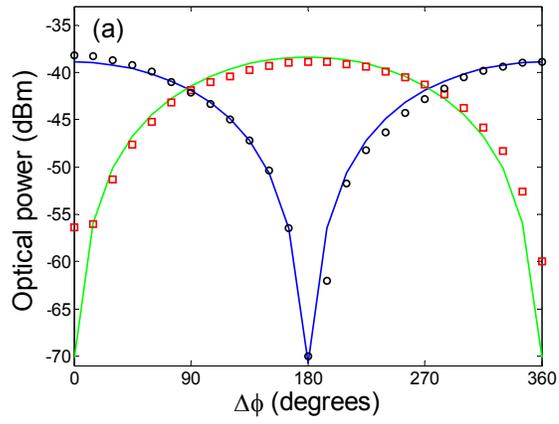

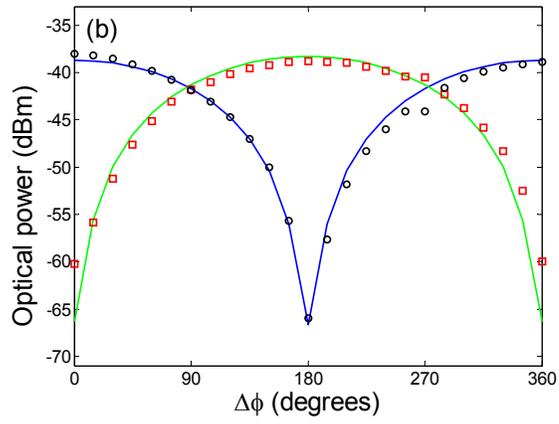

**FIGURE 5**

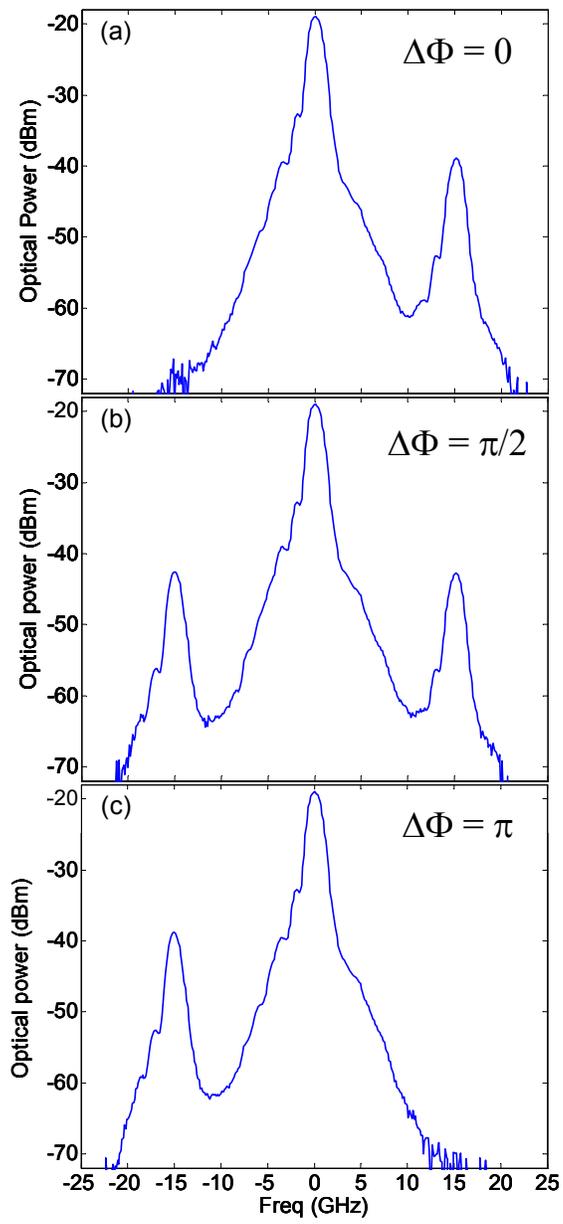

**FIGURE 6**